\documentclass[]{interact}

\usepackage{graphicx}
\usepackage{amsmath,amssymb,amsbsy,amsfonts,amscd,amsthm}
\usepackage{booktabs}
\usepackage{multirow}
\usepackage{algorithm}
\usepackage{algpseudocode}
\usepackage{color,latexsym,courier,lscape}
\usepackage{hyperref}
\usepackage[numbers,sort&compress]{natbib}
\bibpunct[, ]{[}{]}{,}{n}{,}{,}
\renewcommand\bibfont{\fontsize{10}{12}\selectfont}
\makeatletter
\def\NAT@def@citea{\def\@citea{\NAT@separator}}
\makeatother

\newtheorem{theorem}{Theorem}

\newtheorem{proposition}{Proposition}

\newcommand{\bdeta}{\boldsymbol\eta}

\newcommand{\bbeta}{\boldsymbol\beta}

\newcommand{\btheta}{\boldsymbol\theta}
\newcommand{\bphi}{\boldsymbol\varphi}

\newcommand{\bdpsi}{\boldsymbol\psi}

\newcommand{\td}{\text{d}}

\newcommand{\E}{\mathbb E}
\newcommand{\bO}{\mathcal O}

\newcommand{\Q}{\boldsymbol Q}
\newcommand{\q}{\boldsymbol q}

\newcommand{\bdr}{\boldsymbol r}

\newcommand{\0}{\boldsymbol 0}

\begin{document}


\title{A semiparametric two-sample homogeneity test with nonignorable nonresponse using callback data}

\author{
\name{Xinyu Wang\textsuperscript{a}, Tao Yu\textsuperscript{b}, Chunlin Wang\textsuperscript{a,c}\thanks{CONTACT Chunlin Wang. Email: wangc@xmu.edu.cn.} and Pengfei Li\textsuperscript{d}}
\affil{\textsuperscript{a}Department of Statistics and Data Science, School of Economics, Xiamen University, Xiamen, China; \textsuperscript{b}Department of Statistics and Data Science, National University of Singapore, Singapore; \textsuperscript{c}Wang Yanan Institute for Studies in Economics, Xiamen University, Xiamen, China; \textsuperscript{d}Department of Statistics and Actuarial Sciences, University of Waterloo, Waterloo, Canada}
}

\maketitle

\begin{abstract}
Testing the homogeneity of two distributions is fundamental in statistics, but classical procedures may fail under nonignorable nonresponse. In many surveys, callback data record repeated contact attempts and provide auxiliary information about the response mechanism. We develop a semiparametric framework for two-sample homogeneity testing that explicitly incorporates such information. The response mechanism is modelled by a flexible semiparametric callback model, while the two population distributions are linked through a density ratio model. Within this unified framework, we propose an empirical likelihood ratio test for distributional homogeneity and show that, under the null hypothesis, it has a Wilks-type chi-square limit. To facilitate computation, we develop an efficient expectation–maximisation-type algorithm. Simulation results show that the proposed method controls type I error well and achieves substantially higher power than existing methods that ignore nonignorable missingness. An application to real survey income data illustrates its practical value.
\end{abstract}

\begin{keywords}
Callback data;
Density ratio model;
Empirical likelihood;
Nonignorable unit nonresponse;
Semiparametric inference 
\end{keywords}

\section{Introduction}
Testing the homogeneity of two distributions is a fundamental problem in statistics, with broad applications in economics \citep{Bishop1992}, public health \citep{Bricard2020}, and policy evaluation \citep{Duflo2015}. In many empirical studies, researchers collect samples from different populations—such as regions, demographic groups, or time periods—and aim to determine whether their outcome distributions differ \citep{Borjas2019}. For instance, comparing income distributions across regions is central to understanding economic inequality and guiding policy interventions \citep{Bishop1992}. Statistically, this problem is often formulated as testing whether two underlying distributions are identical. 

When complete observations are available, a rich body of literature has been developed for testing distributional homogeneity, including classical nonparametric tests such as the Mann–Whitney test and Kolmogorov–Smirnov test, as well as likelihood-based and empirical likelihood methods \citep{Mann1947,Hodges1958,Owen2001,Cai2017}. However, in many practical applications, particularly in large-scale surveys, data are often subject to substantial nonresponse. This issue becomes especially challenging when the missingness mechanism is \emph{nonignorable} (or missing not at random, MNAR), where the probability of response depends on the unobserved outcome itself \citep{Little2019}. Such mechanisms frequently arise in socioeconomic surveys. For example, empirical evidence suggests that individuals’ willingness to report income depends on their income level, leading to systematic selection bias in distributional inference \citep{Riphahn2005,Korinek2006}. In such settings, directly applying standard two-sample tests to the observed data can lead to biased conclusions, invalid type I error control, and misleading inference.
We refer to Section~\ref{Sec_simulation} for some evidence from simulation studies.

To address nonignorable nonresponse, various identification strategies have been proposed in the literature, including approaches based on instrumental variables \citep{Wang2014,Shao2016,WangLei2021} and shadow variables \citep{Miao2016-1}.
These methods typically rely on additional fully observed covariates and certain identifying assumptions. Along this line, in the two-sample MNAR setting with an instrumental variable, Wang \cite{WangLei2022} studied inference on differences in parameters of interest, such as population means, rather than omnibus tests for distributional homogeneity.

In the unit-nonresponse setting, an alternative approach exploits auxiliary information collected during the survey process itself. In particular, callback data, which record the number of contact attempts required to obtain a response, provide valuable information about the response mechanism. Such data arise naturally in survey practice and have been shown to facilitate identification and adjustment for nonignorable nonresponse under suitable modelling assumptions \citep{Alho1990,Qin2014,Kim2014,Guan2018,Miao2025,Wang2026}. Despite these advances, existing methods that utilise callback data primarily focus on inference for a single population, and the problem of testing distributional homogeneity between two populations remains largely unexplored in this context.

In this paper, we propose a new semiparametric framework for testing the homogeneity of two distributions in the presence of nonignorable nonresponse by leveraging callback data.
Specifically, we model the response mechanism using a semiparametric callback model \citep{Alho1990}, which utilises information from repeated contact attempts to account for selection bias without requiring external instrumental variables. To describe the relationship between the two underlying distributions, we adopt the semiparametric density ratio model (DRM) \citep{Anderson1979,Qin2017}, which provides a flexible yet tractable way to connect the populations while leaving the baseline distribution unspecified. The integration of these two components enables likelihood-based inference for distributional homogeneity under MNAR mechanism, while maintaining robustness to complex and heavy-tailed outcome distributions. Based on the resulting semiparametric likelihood, we further develop an empirical likelihood ratio (ELR) test for assessing the homogeneity hypothesis.

The main contributions of this paper are threefold. First, we develop a unified semiparametric framework that integrates callback models and the DRM for two-sample homogeneity testing under MNAR mechanism. Second, we establish that the proposed ELR statistic enjoys a Wilks-type property under the null hypothesis, leading to a chi-square limiting distribution that facilitates straightforward implementation.
Third, we develop an efficient computational algorithm tailored to the proposed semiparametric framework for parameter estimation and computation of the test statistic.
Simulation studies demonstrate that the proposed method achieves valid control of type I error and improved power compared with conventional approaches that ignore nonignorable missingness. An application to income data from a national health survey further illustrates the practical advantages of the proposed method.

The remainder of the paper is organised as follows. Section 2 presents the model setup and the proposed testing procedure, along with its limiting distribution. Section 3 describes the computational algorithm for numerical implementation. Section 4 reports simulation results, and Section 5 presents a real data application. Section 6 concludes. Technical details are provided in the supplementary material.

\section{Problem Setup and Proposed Testing Procedure}

\subsection{Problem setup}
We first introduce the data structure and the callback-based response mechanism,
and then specify the DRM linking the two populations.

Let $F_0$ and $F_1$ denote the cumulative distribution functions (CDFs) of the two populations of interest, and let ${Y_{i1}, \ldots, Y_{iN_i}}$ be independent samples from $F_i$, for $i = 0,1$. Our objective is to test the homogeneity hypothesis
\[
H_0: F_0 = F_1
\]
against the alternative $F_0 \ne F_1$.

In many survey applications, the outcome variables may be subject to nonresponse, and multiple contact attempts are often made to obtain responses. Let $m$ ($m \ge 2$) denote the maximum number of contact attempts. For each unit $j$ in sample $i$, let $D_{ij}$ denote the callback indicator, where $D_{ij} = k$ if a response is obtained at the $k$-th contact attempt ($k = 1, \ldots, m$), and $D_{ij} = m+1$ if no response is ever obtained. Thus, $Y_{ij}$ is observed if and only if $D_{ij} \le m$.

Define
\[
\pi_{ik}(y)=\Pr(D_{ij}=k \mid Y_{ij}=y,\, D_{ij} \ge k),
\]
the conditional probability that a sampling unit responds at the $k$-th contact attempt given nonresponse in all previous attempts. We adopt the semiparametric callback model of Alho \cite{Alho1990}, which specifies
\begin{equation}\label{Alho}
\pi_{ik}(y)
= \pi(y;\alpha_{ik},\bbeta_i) =
\frac{\exp\{\alpha_{ik}+\bbeta_i^\top \bdr(y)\}}
     {1+\exp\{\alpha_{ik}+\bbeta_i^\top \bdr(y)\}},
\quad i=0,1; \; k=1,\ldots,m,
\end{equation}
where $\alpha_{ik}$ and $\bbeta_i$ are unknown parameters, and $\bdr(y)$ is a pre-specified vector of basis functions. The intercept $\alpha_{ik}$ is allowed to vary across contact attempts, whereas the slope parameter $\beta_i$ is assumed to be common across attempts within each population. This specification corresponds to the continuum of resistance assumption and ensures model identifiability \citep{Miao2025}.
Moreover, the dependence of $\pi_{ik}(y)$ on $y$ allows the response probability to depend on the unobserved outcome, thereby accommodating nonignorable missingness.

To characterise the relationship between the two populations, we adopt the DRM, which assumes
\begin{equation}
\label{DRM}
dF_1(y) = \exp\{\theta_0 + \btheta_1^\top \q(y)\} \, dF_0(y)
= \exp\{\btheta^\top \Q(y)\} \, dF_0(y),
\end{equation}
where $\q(y)$ is a pre-specified  $q$-dimensional  nontrivial basis function, $\Q(y)=(1,\q^\top(y))^\top$, and $\btheta = (\theta_0, \btheta_1^\top)^\top$ is an unknown parameter vector. The baseline distribution $F_0$ is left unspecified.
Under the DRM \eqref{DRM}, the homogeneity hypothesis $H_0: F_0 = F_1$ is equivalent to
\begin{equation}
\label{H0_e}
H_0: \btheta = \0 .
\end{equation}

The DRM is a semiparametric model due to the unspecified baseline distribution $F_0$, and is equivalent to a logistic regression formulation for binary response modelling \citep{Qin1997}. It provides a flexible framework for modelling relationships between distributions. For example, when $\q(y) = \log(y)$, the DRM includes log-normal and gamma families with a common scale parameter; when $\q(y) = y$, it includes exponential distributions with different rates. The combination of the callback model and the DRM enables identification under nonignorable missingness without requiring external instrumental variables.

Owing to its flexibility and efficiency, the DRM has been widely used in case-control studies \citep{Qin2015}, quantile estimation \citep{ChenLiu2013}, multiple-sample testing \citep{Cai2017,Wang2017,Wang2018}, receiver operating characteristic analysis \citep{QinZhang2003,Yuan2021}, and dominance index estimation \citep{Zhuang2019}. See \cite{Qin2017} for a comprehensive review.

We next develop a semiparametric likelihood and a testing procedure for distributional homogeneity based on the observed data $\{(Y_{ij}, D_{ij}) : i = 0,1; \; j = 1, \ldots, N_i\}$.

\subsection{Proposed empirical likelihood ratio test}
\label{Sec_methodology}

To proceed, we construct a semiparametric likelihood under the DRM \eqref{DRM} and the callback model \eqref{Alho}. Because the outcome may be missing not at random, the likelihood must account for both the response mechanism and the outcome distribution, which we do by combining the conditional response probabilities with the density ratio representation.

Let $n_i=\sum_{j=1}^{N_i} I(D_{ij}\leq m)$ denote the (random) number of respondents in sample $i$, and define $N=N_0+N_1$ and $n=n_0+n_1$.
Under nonignorable nonresponse, $n_i$ is informative for sample selection bias.
Without loss of generality, we relabel the observations so that the first $n_i$ correspond to respondents and the remaining $N_i-n_i$ correspond to nonrespondents.

To construct the full likelihood, we introduce the following notation. For $i=0,1$, let $\rho_{ik}(y) = \Pr(D_{i1}=k \mid Y_{i1}=y)$, $\rho_{i}(y) = \Pr(D_{i1}\le m \mid Y_{i1}=y)$ and $\eta_i = \Pr(D_{i1} \le m)$. Under the callback model \eqref{Alho}, we have
\[
\begin{aligned}
\rho_{i1}(y) & =
\rho_{1}(y;\boldsymbol\varphi_i)
=
\pi(y;\alpha_{i1},\boldsymbol\beta_i),\\
\rho_{ik}(y) & =
\rho_{k}(y;\boldsymbol\varphi_i)
= 
\pi(y;\alpha_{ik},\boldsymbol\beta_i)
\prod_{j=1}^{k-1}\{1-\pi(y;\alpha_{ij},\boldsymbol\beta_i)\},
\qquad k=2,\ldots,m,\\
\rho_{i}(y) & = \rho(y;\boldsymbol\varphi_i)
= 
\sum_{k=1}^m \rho_{k}(y;\boldsymbol\varphi_i),
\end{aligned}
\]
where $\boldsymbol\varphi_i=(\boldsymbol\alpha_i^\top,\boldsymbol\beta_i^\top)^\top$ and
$\boldsymbol\alpha_i=(\alpha_{i1},\ldots,\alpha_{im})^\top$.

Following the likelihood development of Qin and Follmann \cite{Qin2014} and Wang et al. \cite{Wang2026} for the one-sample case, the full likelihood  based on $\{(Y_{ij}, D_{ij}) : j = 1, \ldots, N_i\}$ can be written as
\begin{equation}
\label{likelihood0}
L_i
= {N_i\choose n_i}\left[\prod_{j=1}^{n_i}
\prod_{k=1}^m
\{\rho_{k}(Y_{ij};\boldsymbol\varphi_i)\, \td F_i(Y_{ij})\}^{I(D_{ij}=k)}
\right]
\!(1-\eta_i)^{N_i-n_i} .
\end{equation}
Details of the derivation are provided in Section 1.1 of the supplementary material.

With \eqref{likelihood0},  the likelihood
based on $\{(Y_{ij}, D_{ij}) : i=0,1; \; j = 1, \ldots, N_i\}$
is given by
\begin{equation}
\label{L}
L = L_0 L_1
= \prod_{i=0}^1 {N_i\choose n_i}\left[\prod_{j=1}^{n_i}
\prod_{k=1}^m
\{\rho_{k}(Y_{ij};\boldsymbol\varphi_i)\, \td F_i(Y_{ij})\}^{I(D_{ij}=k)}
\right]
\!(1-\eta_i)^{N_i-n_i}.
\end{equation}
Following the empirical likelihood (EL) framework of Owen \cite{Owen2001} under the DRM, we represent  $F_0$ and $F_1$ as
\begin{equation}
\label{F0F1}
\begin{aligned}
F_0(y)
= \sum_{i=0}^1 \sum_{j=1}^{n_i} p_{ij}\, I(Y_{ij}\le y),~~
F_1(y)
= \sum_{i=0}^1 \sum_{j=1}^{n_i}
p_{ij}\exp\{\boldsymbol\theta^\top \Q(Y_{ij})\}
I(Y_{ij}\le y),
\end{aligned}
\end{equation}
where $p_{ij}=\td F_0(Y_{ij})$ for $i=0,1$ and $j=1,\ldots,n_i$.

Let $\bphi = (\bphi_0^\top, \bphi_1^\top)^\top$ and $\bdeta = (\eta_{0}, \eta_{1})^\top$.
Denote the full parameter vector by $\bdpsi = (\btheta,\bphi,\bdeta,F_0)$.
Up to an additive constant that does not depend on $\bdpsi$,
the semiparametric log-EL function is
\begin{equation} \begin{aligned}
\label{log_lik}
\ell(\bdpsi) = & \sum_{i=0}^{1}\sum_{j=1}^{n_i} \sum_{k=1}^m I(D_{ij}=k)\log\{\rho_{k}(Y_{ij};\bphi_i)\} + \sum_{i=0}^{1}\sum_{j=1}^{n_i}\log (p_{ij}) \\
& + \sum_{j=1}^{n_1} \btheta^\top \Q(Y_{1j}) + \sum_{i=0}^1(N_i-n_i)\log(1-\eta_{i}).
\end{aligned} \end{equation}
The parameter $\bdpsi$ is subject to the feasible set $\mathcal C = \mathcal C_1 \cap \mathcal C_2$, where
\begin{equation} \begin{aligned}
\label{const1}
\mathcal C_1 = \Bigg\{\bdpsi : p_{ij}> 0, ~\sum_{i=0}^1\sum_{j=1}^{n_i} p_{ij} = 1, ~\sum_{i=0}^1\sum_{j=1}^{n_i} p_{ij}\exp\{\btheta^\top \Q(Y_{ij})\}=1\Bigg\},
\end{aligned} \end{equation}
and
\begin{equation} \begin{aligned}
\label{const2}
\mathcal C_2 = \Bigg\{\bdpsi :
\sum_{i=0}^1\sum_{j=1}^{n_i} p_{ij} \rho(Y_{ij};\bphi_0)=\eta_{0}, ~\sum_{i=0}^1\sum_{j=1}^{n_i} p_{ij}\exp\{\btheta^\top \Q(Y_{ij})\}\rho(Y_{ij};\bphi_1)=\eta_{1}\Bigg\}.
\end{aligned} \end{equation}
Here,  $\mathcal C_1$ ensures that estimates of $F_0$ and $F_1$ are valid CDFs,  and $\mathcal C_2$ follows the definition of $\eta_0$ and $\eta_1$ to account for selection bias.

The ELR test statistic for testing $H_0:\btheta = \0$ versus $H_1:\btheta \neq \0$ is defined as
\begin{equation} \begin{aligned}
R_N = 2\{\ell(\hat{\bdpsi}) - \ell(\tilde{\bdpsi})\},
\end{aligned} \end{equation}
where $\hat{\bdpsi} = (\hat{\btheta},\hat{\bphi},\hat{\bdeta},\hat{F}_0) = \arg\max\limits_{\bdpsi\in\mathcal{C}} \ell(\bdpsi)$ and $\tilde{\bdpsi} = (\0,\tilde{\bphi},\tilde{\bdeta},\tilde{F}_0) = \arg\max\limits_{\bdpsi\in\mathcal{C}, \btheta = \0}\ell(\bdpsi)$.
We reject $H_0$ for large values of $R_N$, with critical values determined by the limiting distribution presented below. The technical details are provided in the supplementary material.

\begin{theorem}
\label{lim_chi2}
    Assume that Conditions C1–C8 in Section 1.3 of the supplementary material hold.
    Under the null hypothesis $H_0$  in \eqref{H0_e}, as $N\to\infty$,
    \[
    R_N \stackrel{d}{\to} \chi^2_q,
    \]
    where $\stackrel{d}{\to}$ denotes convergence in distribution and $q$ is the dimension of $\q(y)$.
\end{theorem}

We make four remarks on the ELR test and the result in Theorem~\ref{lim_chi2}.
First, closed-form expressions for $\hat\bdpsi$ and $\tilde\bdpsi$ are generally unavailable. In Section~\ref{Sec_EM}, we develop a stable and computationally efficient EM algorithm to compute $\hat\bdpsi$, $\tilde\bdpsi$, and the ELR statistic $R_N$.
Second, although the parameter vector $\btheta$ has dimension $q+1$, the number of free parameters is $q$, since $\btheta_1=\0$ implies $\theta_0=0$ under the constraints in  $\mathcal{C}_1$. Consequently, the null limiting distribution of $R_N$ is $\chi^2_q$.
Third, Theorem~\ref{lim_chi2} shows that the ELR test is asymptotically pivotal, so its critical values can be obtained directly from the quantiles of a chi-square distribution with $q$ degrees of freedom.
Fourth, the DRM in \eqref{DRM} is always satisfied under the null hypothesis for any given basis function $\q(y)$. Consequently, the asymptotic size of the proposed ELR test is preserved regardless of the choice of $\q(y)$, highlighting its robustness.

\section{Computational Algorithm}
\label{Sec_EM}

In this section, we develop a computational algorithm for estimating $\hat{\psi}$. We view callback data as a special case of missing data, in which the outcome variables are unobserved for nonrespondents, suggesting the classical  expectation–maximisation (EM) algorithm for parametric models \citep{Dempster1977} as a natural starting point for numerical implementation. Although the proposed algorithm is motivated by the EM framework, its development is methodologically nontrivial, and establishing its convergence requires additional theoretical justification. In particular, the EL formulation must be carefully incorporated to derive a closed-form expression for the conditional expectation in the E-step. In the M-step, we employ profile likelihood and dual likelihood techniques \citep{Cai2017} to address the constrained maximisation problem, which can be reduced to fitting a weighted logistic regression model. These steps do not follow directly from standard EM theory and instead require substantial methodological development. Finally, based on the estimation procedure for $\hat{\psi}$, we provide a remark at the end of this section describing the computation of $\tilde{\psi}$, which can be obtained via a slight modification of the proposed algorithm.

Following the EM framework, we treat the unobserved outcomes of nonrespondents as latent variables and iteratively maximise the expected complete-data log-likelihood. To formalise this approach, recall that the observed data consist of the combined samples
\(
\mathcal{O} = \{(Y_{ij}, D_{ij}) : i = 0,1;\; j = 1, \ldots, n_i\} \cup \{D_{it} = m+1 : i = 0,1;\; t = n_i+1, \ldots, N_i\},
\)
where for nonrespondents only the callback indicator is observed. Let $\mathcal{O}^* = \{Y^*_{it} : i = 0,1;\; t = n_i+1, \ldots, N_i\}$ denote the latent outcomes for the $N - n = \sum_{i=0}^1 (N_i - n_i)$ nonrespondents. The complete-data likelihood based on $\mathcal{O} \cup \mathcal{O}^*$ is then given by
\begin{equation}
\begin{aligned}
\prod_{i=0}^{1} \prod_{j=1}^{n_i} \prod_{k=1}^m \Big\{\rho_{k}(Y_{ij};\bphi_i)\,\td F_i(Y_{ij})\Big\}^{I(D_{ij}=k)}
\times \prod_{i=0}^{1} \prod_{t=n_i+1}^{N_i} \Big\{1-\rho(Y^*_{it};\bphi_i)\Big\}\,\td F_i(Y^*_{it}).
\end{aligned}
\end{equation}

Recall the representations of $F_0$ and $F_1$ in \eqref{F0F1}. Under this formulation, each latent variable $Y^*_{it}$ takes values in the support of the observed outcomes $\{Y_{01}, \ldots, Y_{0n_0}, Y_{11}, \ldots, Y_{1n_1}\}$. This discrete support representation allows the complete-data log-likelihood to be expressed in terms of the probability masses $\{p_{ij}\}$. The complete-data log-likelihood can be written as
\begin{equation} \begin{aligned}
\label{ell_EM}
\ell_{\text{EM}}(\bdpsi) = & \sum_{i=0}^1 \sum_{j=1}^{n_i} \sum_{k=1}^m I(D_{ij}=k)\log\{\rho_{k}(Y_{ij};\bphi_i)\} + \sum_{i=0}^1 \sum_{j=1}^{n_i}\log(p_{ij}) \\
& + \sum_{j=1}^{n_1} \btheta^\top \Q(Y_{1j}) + \sum_{t=n_0+1}^{N_0}\sum_{i=0}^{1}\sum_{j=1}^{n_i} I(Y^*_{0t} = Y_{ij})\big[\log(p_{ij}) + \log\{1-\rho(Y_{ij};\bphi_0)\}\big] \\
& + \sum_{t=n_1+1}^{N_1}\sum_{i=0}^{1}\sum_{j=1}^{n_i} I(Y^*_{1t} = Y_{ij})\big[\log(p_{ij}) + \btheta^\top \Q(Y_{ij}) + \log\{1-\rho(Y_{ij};\bphi_1)\}\big],
\end{aligned} \end{equation}
where $\bdpsi$ is subject to the constraint set $\mathcal C = \mathcal C_1 \cap \mathcal C_2$, where $\mathcal C_1$ and $\mathcal C_2$ are defined in \eqref{const1} and \eqref{const2}, respectively.

Our algorithm is composed of two steps: the \emph{E-step} and the \emph{M-step}. Suppose that after $r$ iterations, the current parameter estimate is $\bdpsi^{(r)}$. The $(r+1)$th iteration then proceeds as follows.

For the \emph{E-step}, we calculate the conditional expectation of $\ell_{\text{EM}}(\bdpsi)$ given $\bdpsi^{(r)}$ and the observed data $\bO$, defined as
\begin{equation}
\label{M_func}
M(\bdpsi \mid \bdpsi^{(r)}) = \E\{\ell_\text{EM}(\bdpsi) \mid  \bO,\bdpsi^{(r)}\},
\end{equation}
which is maximised in the subsequent \emph{M-step}.

To evaluate $M(\bdpsi \mid \bdpsi^{(r)})$, we first derive the conditional expectation
$\E\{I(Y^*_{0t} = Y_{ij})\mid \bO, \bdpsi^{(r)}\} $. For $t = n_0+1,\ldots,N_0$, we have
\begin{equation} \begin{aligned}
\label{post1}
\textstyle \E\{I(Y^*_{0t} = Y_{ij}) \mid \bO, \bdpsi^{(r)}\} &= \Pr(Y^*_{0t} = Y_{ij} \mid  D_{0t}=m+1, Y_{ij}, \bdpsi^{(r)})\\
& = \frac{\Pr(Y^*_{0t} = Y_{ij} \mid Y_{ij}, \bdpsi^{(r)})\Pr(D_{0t} = m+1 \mid Y^*_{0t} = Y_{ij}, Y_{ij}, \bdpsi^{(r)})}{\Pr(D_{0t} = m+1 \mid  \bdpsi^{(r)})}\\
& = \frac{p_{ij}^{(r)}\{1-\rho(Y_{ij};\bphi_0^{(r)})\}}{1-\eta_0^{(r)}},
\end{aligned} \end{equation}
where the second equality follows from Bayes's rule. Similarly, for $t = n_1+1,\ldots,N_1$, we have
\begin{equation} \begin{aligned}\label{post2}
\E\{I(Y^*_{1t} = Y_{ij}) \mid \bO, \bdpsi^{(r)}\}
& = \frac{p_{ij}^{(r)}\exp\{\btheta^{(r)\top}\Q(Y_{ij})\} \{1-\rho(Y_{ij};\bphi_1^{(r)})\}}{1-\eta_1^{(r)}}. 
\end{aligned} \end{equation}

Combining \eqref{ell_EM}--\eqref{post2}, we obtain
\[
M(\bdpsi\mid\bdpsi^{(r)}) = \ell_1^{(r)}(\btheta, F_0) + \ell_2^{(r)}(\bphi),
\]
where
\begin{equation} \begin{aligned}
\label{Q}
\ell_1^{(r)}(\btheta,F_0) & = \sum_{i=0}^1\sum_{j=1}^{n_i} (w_{ij}^{(r)} + v_{ij}^{(r)} + 1)\log(p_{ij}) + \sum_{j=1}^{n_0} v^{(r)}_{0j} \btheta^\top \Q(Y_{0j}) + \sum_{j=1}^{n_1} (v^{(r)}_{1j} + 1)\btheta^\top \Q(Y_{1j}),\\
\ell_2^{(r)}(\bphi) & = \sum_{i=0}^1 \sum_{j=1}^{n_i} \sum_{k=1}^m I(D_{ij}=k)\log\{\rho_{k}(Y_{ij};\bphi_i)\} + \sum_{i=0}^1 \sum_{j=1}^{n_i} w_{ij}^{(r)} \log\{1-\rho(Y_{ij};\bphi_0)\} \\
& \quad + \sum_{i=0}^1 \sum_{j=1}^{n_i} v_{ij}^{(r)} \log\{1-\rho(Y_{ij};\bphi_1)\}.\\
\end{aligned} \end{equation}
The weights $w_{ij}^{(r)}$ and $v_{ij}^{(r)}$ are given by
\begin{equation} \begin{aligned}
\label{update_wv}
w_{ij}^{(r)} &= (N_0 - n_0)\frac{p_{ij}^{(r)}\{1-\rho(Y_{ij};\bphi_0^{(r)})\}}{1-\eta_0^{(r)}},\\
v_{ij}^{(r)} &= (N_1 - n_1)\frac{p_{ij}^{(r)}\exp\{{\btheta}^{(r)\top}\Q(Y_{ij})\} \{1-\rho(Y_{ij};\bphi_1^{(r)})\}}{1-\eta_1^{(r)}}.
\end{aligned} 
\end{equation}

For \emph{M-step}, we update the value of $\bdpsi$ from $\bdpsi^{(r)}$ to $\bdpsi^{(r+1)}$ by solving $\bdpsi^{(r+1)} = \arg\max_{\bdpsi} M(\bdpsi \mid \bdpsi^{(r)})$ subject to the constraints $\mathcal C = \mathcal C_1 \cap \mathcal C_2$ .
Owing to the additive structure of $M(\bdpsi \mid \bdpsi^{(r)})$, this maximisation can be carried out in two steps.

First, we update $\btheta^{(r+1)}$ and $p_{ij}^{(r+1)}$'s by maximising
$\ell_1^{(r)}(\btheta, F_0)$ subject to the constraint set $\mathcal C_1$.
Due to the EL formulation of $F_0$, this constrained maximisation can be simplified via a profile likelihood argument.
As shown in Section 4.2 of the supplementary material, $\btheta^{(r+1)}$ can be obtained as
\begin{equation}
\label{udpate_btheta}
\btheta^{(r+1)} = \arg\max_{\btheta} \ell_{1\text{p}}^{(r)}(\btheta),
\end{equation}
where
\begin{equation} \begin{aligned}
\label{ell_1p}
 \ell_{1\text{p}}^{(r)}(\btheta) &= -\sum_{i=0}^1\sum_{j=1}^{n_i} (w_{ij}^{(r)} + v_{ij}^{(r)} + 1)\log\left(1+\frac{N_1}{N}\left[\exp\{\btheta^\top \Q(Y_{ij})\}-1\right] \right) \\
 & \quad + \sum_{i = 0}^1 \sum_{j=1}^{n_i} v^{(r)}_{ij} \btheta^\top \Q(Y_{ij}) +\sum_{j=1}^{n_1} \btheta^\top \Q(Y_{1j}).
\end{aligned} \end{equation}
The updated $\{p_{ij}\}$ are given by
\begin{equation} \begin{aligned}
\label{update_pij}
p^{(r+1)}_{ij} = \frac{1}{N} \frac{w_{ij}^{(r)} + v_{ij}^{(r)} + 1}{1+ \frac{N_1}{N}\big[\exp\{\btheta^{(r+1)\top} \Q(Y_{ij})\}-1\big]}.
\end{aligned} \end{equation}

Moreover, the form of $\ell_{1\mathrm{p}}^{(r)}(\btheta)$ implies that $\btheta^{(r+1)}$ can be obtained by fitting a weighted logistic regression model, which is readily implemented using standard statistical software. Details are provided in Section 4.2 of the supplementary material.

Next, we update $\bphi^{(r)}$ to $\bphi^{(r+1)}$ by solving
\begin{equation}
\label{update_bphi}
\bphi^{(r+1)} = \arg\max_{\bphi} \ell_2^{(r)}(\bphi).
\end{equation}
As shown in Section 4.3 of the supplementary material, this optimisation can be carried out by fitting a weighted logistic regression model, which is readily implemented using standard statistical software.

Finally, we update $\bdeta^{(r+1)}$ as
\begin{equation} \begin{aligned}
\label{update_eta}
\eta_{0}^{(r+1)} &= \sum_{i=0}^1\sum_{j=1}^{n_i} p_{ij}^{(r+1)} \rho(Y_{ij};\bphi_0^{(r+1)}),\\
\eta_{1}^{(r+1)} &= \sum_{i=0}^1\sum_{j=1}^{n_i} p_{ij}^{(r+1)} \exp\{\btheta^{(r+1)\top}\Q(Y_{ij})\}\rho(Y_{ij};\bphi_1^{(r+1)}).
\end{aligned} \end{equation}
This completes the \emph{M-step}.

We summarise the above procedure in Algorithm~\ref{alg:mele}. The following proposition establishes the monotonicity property of Algorithm~\ref{alg:mele}.

\begin{algorithm}[!ht]
\caption{Computational algorithm for $\hat{\bdpsi}$}
\label{alg:mele}
\begin{algorithmic}[1]
\Require Observed data: $\{(Y_{ij}, D_{ij}): i=0,1; \; j=1,\ldots,n_i \} \cup \{D_{it}=m+1: i = 0,1; \; t=n_i+1,\ldots,N_i\}$, and a convergence threshold $\epsilon$ (e.g., $10^{-5}$)
\Ensure Estimate $\hat{\bdpsi}$

\State \textbf{initialise:} Choose an initial value $\bdpsi^{(0)} =({\btheta}^{(0)}, {\bphi}^{(0)}, {\bdeta}^{(0)}, F_0^{(0)})$ and set $r=0$.

\Repeat
\State \textbf{E-step:} Compute $w_{ij}^{(r)}$ and $v_{ij}^{(r)}$ according to \eqref{update_wv}.

\State \textbf{M-step:}
Update $\btheta^{(r+1)}$ via \eqref{udpate_btheta},
$p_{ij}^{(r+1)}$ via \eqref{update_pij},
$\bphi^{(r+1)}$ via \eqref{update_bphi},
and $\bdeta^{(r+1)}$ via \eqref{update_eta}.
Set $r \leftarrow r+1$.

\Until{$\ell(\bdpsi^{(r+1)}) - \ell(\bdpsi^{(r)}) < \epsilon$}.

\State \textbf{Return:} $\hat{\bdpsi}$.
\end{algorithmic}
\end{algorithm}

\begin{proposition}
\label{nondecreasing}
For the  algorithm described in Algorithm~\ref{alg:mele}, we have, for all $r \geq 0$,
\[
\ell(\bdpsi^{(r+1)}) \geq \ell(\bdpsi^{(r)}).
\]
\end{proposition}

We make several remarks on Algorithm~\ref{alg:mele}.
First, since $\ell(\bdpsi) \leq 0$, Proposition~\ref{nondecreasing} ensures convergence of the algorithm.
Second, extensive numerical experiments indicate that the algorithm is insensitive to the choice of initial values for $\bdpsi$. In our numerical studies, we initialise $\bphi^{(0)} = \mathbf{0}$, $\btheta^{(0)} = \mathbf{0}$, $p_{ij}^{(0)} = 1/n$ for $j = 1,\ldots,n_i$ and $i = 0,1$, and $\eta_0^{(0)} = \eta_1^{(0)} = 1 - 2^{-m}$.
Third, under the null hypothesis $H_0$, $\tilde{\bdpsi}$ can be computed using essentially the same procedure as Algorithm~\ref{alg:mele}, with $\btheta^{(r)} \equiv \mathbf{0}$.

\section{Simulation Study}
\label{Sec_simulation}

In this section, we conduct Monte Carlo simulations to evaluate the finite-sample performance of the proposed ELR test statistic $R_N$ developed in Section~\ref{Sec_methodology} and computed via the EM algorithm in Section~\ref{Sec_EM}. Specifically, we assess the ability of the proposed test to control the type I error rate under the null hypothesis of distributional homogeneity, as well as its power under various alternatives. We also compare its performance (denoted by \emph{Our ELR}) with several commonly used two-sample procedures, including the two-sample $t$-test (denoted by \emph{t-test}), the Wilcoxon--Mann--Whitney test (denoted by \emph{Wilcoxon}), the Kolmogorov--Smirnov test (denoted by \emph{KS}), and the ELR test under the DRM proposed by Cai et al.~(2017) (denoted by \emph{Cai's ELR}). Note that all methods, except our proposed ELR, assume no missing outcomes.

We generate two-sample data with $m = 2$ callbacks and sample sizes $N_0 = N_1 = 500$ and $N_0 = N_1 = 1000$. For the missing data mechanism \eqref{Alho}, we assume $\bdr(y) = \log(y)$, with $\bphi_0^* = (-0.5, 0.5, -0.5)^\top$ and $\bphi_1^* = (0, 1, -0.5)^\top$.
For the outcome distributions $F_0$ and $F_1$, we consider three specifications: ${\rm Exp}(a)$ (the exponential distribution with rate parameter $a$), ${\rm LN}(a,b)$ (the log-normal distribution with mean $a$ and standard deviation $b$ on the log scale), and ${\rm Gamma}(a,b)$ (the Gamma distribution with shape parameter $a$ and scale parameter $b$).
For the DRM in \eqref{DRM}, the basis function $\q(y)$ is specified according to the underlying distribution: $\q(y) = y$ for the exponential distribution, and $\q(y) = \log(y)$ for the log-normal and Gamma distributions. These choices ensure that the DRM is correctly specified. By appropriately tuning the parameters of $F_0$ and $F_1$, we generate data under both the null hypothesis $H_0: F_0 = F_1$ to evaluate empirical size and the alternative $H_1: F_0 \neq F_1$ to assess power.
The proposed homogeneity test and Cai's ELR test under the DRM \eqref{DRM} are implemented using the correctly specified $\q(y)$.

The simulated sizes at the 5\% significance level for the five homogeneity tests, based on 2000 repetitions, are summarised in Table~\ref{size_diff_phi}. The critical values for the proposed ELR test are obtained from the limiting distribution in Theorem~\ref{lim_chi2}. We observe that the proposed ELR test maintains the Type I error well controlled at the nominal 5\% level, indicating that the limiting distribution of the ELR provides an accurate approximation to its finite-sample distribution.
In contrast, the other four tests exhibit inflated Type I error rates, exceeding the nominal 5\% level. This occurs because these methods do not account for missing outcomes and effectively test whether the distributions of the observed outcomes differ between the two groups. When the nonresponse mechanisms differ across groups ($\bphi_0^* \neq \bphi_1^*$), the observed outcome distributions can differ even if $F_0 = F_1$, leading to invalid inference.

\begin{table}[!h]
\small
\renewcommand\arraystretch{0.8}
\caption{Simulated sizes of the five tests under the null models at the 5\% significance level
}
\label{size_diff_phi}
\centering
\begin{tabular}{l ccccccccc}
\toprule
$F_0$ = $F_1$ & Our ELR & $t$-test & Wilcoxon & KS & Cai's ELR \\
\midrule
\multicolumn{6}{c}{$N_0 = N_1 = 500$} \\
Exp$(1)$ & 0.0505 & 0.1070 & 0.1165 & 0.0885 & 0.1085 \\
LN$(0,1)$ & 0.0475 & 0.1090 & 0.1195 & 0.0890 & 0.1265 \\
Gamma$(2.5,1)$ & 0.0505 & 0.0895 & 0.0915 & 0.0790 & 0.0935 \\
\multicolumn{6}{c}{$N_0 = N_1 = 1000$} \\
Exp$(1)$ & 0.0505 & 0.1530 & 0.1805 & 0.1485 & 0.1530 \\
LN$(0,1)$ & 0.0470 & 0.1390 & 0.1700 & 0.1455 & 0.1715\\
Gamma$(2.5,1)$ & 0.0510 & 0.1250 & 0.1330 & 0.1140 & 0.1320 \\
\bottomrule
\end{tabular}
\end{table}

Table~\ref{power_diff_phi} summarises the simulated powers of the five tests under the alternative models, based on 2000 repetitions. Since the other four methods fail to achieve the nominal Type I error control under the null models, we compute the simulated powers using critical values obtained from their empirical null distributions to ensure a fair comparison.
The results demonstrate the superiority of the proposed method in terms of power: the proposed ELR test achieves substantially higher power than the competing methods, leading to a lower Type II error rate.

\begin{table}[!h]
\small
\renewcommand\arraystretch{0.8}
\caption{Simulated powers of the five tests under the alternative models at the 5\% significance level
}
\label{power_diff_phi}
\centering
\begin{tabular}{ll ccccccccc}
\toprule
$F_0$ & $F_1$ & Our ELR & $t$-test & Wilcoxon & KS & Cai's ELR \\
\midrule
\multicolumn{7}{c}{$N_0 = N_1 = 500$} \\
Exp$(0.8)$ & Exp$(1)$ & 0.7890 & 0.4185 & 0.2710 & 0.2285 & 0.4360\\
LN$(0.2,1)$ & LN$(0,1)$ & 0.6140 & 0.2020 & 0.3170 & 0.2715 & 0.3215\\
Gamma$(2.8,1)$ & Gamma$(2.5,1)$ & 0.4790 & 0.3690 & 0.3870 & 0.3135 & 0.4455 \\
\multicolumn{7}{c}{$N_0 = N_1 = 1000$} \\
Exp$(0.8)$ & Exp$(1)$ & 0.9745 & 0.6960 & 0.4140 & 0.4000 & 0.7015\\
LN$(0.2,1)$ & LN$(0,1)$ & 0.9010 & 0.3125 & 0.5145 & 0.4305 & 0.5255 \\
Gamma$(2.8, 1)$ & Gamma$(2.5, 1)$ & 0.7765 & 0.6265 & 0.6910 & 0.6140 & 0.7275\\
\bottomrule
\end{tabular}
\end{table}

In summary, traditional homogeneity tests designed for complete data may lead to severe misjudgments when applied to data with nonignorable missingness. However, the proposed homogeneity test ensures valid statistical inference and reliable testing results, with correct Type I error control and considerable power, when the missingness mechanism is correctly specified. Therefore, the proposed homogeneity test is highly recommended for analysing datasets with nonignorable missingness.
\section{Real Data Example}
\label{Sec_example}
In this section, we apply the proposed homogeneity test to a real dataset from the National Health Interview Survey (NHIS).
The NHIS, conducted by the National Center for Health Statistics, is designed to monitor the health status of the U.S.\ population through personal household interviews.
We analyse family income data from the 2019 NHIS, which includes 55{,}012 families. Among these families, 31.8\% responded within the first three contact attempts, 26.3\% responded in subsequent attempts, and 41.9\% did not respond.
A paradata file containing callback information is available for this survey, with a maximum of 37 contact attempts.
Similar to Qin and Follmann \cite{Qin2014}, we categorise the callback information as follows: $D = 1$ if the family responded within the first three attempts, $D = 2$ if the response occurred after the third attempt, and $D = m + 1$ if no response was obtained. This categorised callback information, with $m = 2$, is incorporated into the proposed ELR test.

The income data are divided into four groups according to the geographical region of residence: Northeast (NE), Midwest (MW), South (S), and West (W).
For illustration, we focus on comparing the NE and W groups to assess whether their income distributions are identical.
The sample sizes are $N_0 = 10{,}192$ and $N_1 = 12{,}812$ for the NE and W groups, respectively, with corresponding numbers of observed income data $n_0 = 5{,}406$ and $n_1 = 7{,}803$.
To implement the proposed method, we specify $\bdr(y)$ in \eqref{Alho} and $\q(y)$ in \eqref{DRM}, both selected from the candidate set $\{y, \log(y), (y, \log(y))^\top\}$, yielding nine possible combinations; the Bayesian Information Criterion  selects $\bdr(y) = y$ and $\q(y) = y$.

To examine potential differences in the missingness mechanisms between the two groups, we first compute $\hat{\bphi}_0 - \hat{\bphi}_1$, then obtain their bootstrap standard errors based on 200 bootstrap replicates, and finally compute the $p$-values for testing whether each component of $\bphi_0$ equals the corresponding component of $\bphi_1$.
The bootstrap samples are generated according to \eqref{Alho}, based on the estimators $\hat{F}_0$, $\hat{F}_1$, and $\hat{\bphi}$.
Table~\ref{table_phi} summarises the results, which provide strong evidence of a discrepancy between $\bphi_0$ and $\bphi_1$, particularly between $\beta_0$ and $\beta_1$.
This finding suggests that the response mechanisms differ between the NE and W groups and further supports the validity and applicability of the callback model assumption in~\eqref{Alho}.

\begin{table}[!h]
\small
\renewcommand\arraystretch{0.8}
\caption{$\hat{\bphi}_0 - \hat{\bphi}_1$, their bootstrap standard errors (SEs) and $p$-values for comparing missingness mechanisms between the NE and W groups}
\label{table_phi}
\centering
\begin{tabular}{c r r c r r c r r}
\toprule
$\hat\bphi_0-\hat\bphi_1$ & estimate & SE & $p$-value \\
\midrule
$\hat{\alpha}_{01}-\hat{\alpha}_{11}$ &  0.2184 & 0.1233 & 0.0765 \\
$\hat{\alpha}_{02}-\hat{\alpha}_{12}$ &  0.2898 & 0.1486 & 0.0512 \\
$\hat{\beta}_{0}-\hat{\beta}_{1}$     & $-0.0485$ & 0.0127 & 0.0001 \\
\bottomrule
\end{tabular}
\end{table}

Next, we apply the proposed two-sample homogeneity test to the income data of NE and W groups.
For comparison, the other tests considered in the simulation study are also applied to these two groups, with the basis function in Cai’s DRM-based test specified as $\q(y) = y$.
Table~\ref{real_example} reports the $p$-values for all homogeneity tests.
As shown in Table~\ref{real_example}, the standard two-sample $t$-test, Wilcoxon test, Kolmogorov--Smirnov (KS) test, and Cai’s ELR test all fail to reject the null hypothesis at the 5\% significance level, suggesting no significant difference between the two groups.
In contrast, the proposed ELR test yields a test statistic of $20.1074$ and a $p$-value of $7.321 \times 10^{-6}$, providing strong evidence against the null hypothesis and indicating a substantial difference in family income distributions between the NE and W regions.
These contrasting results highlight the importance of accounting for nonignorable missingness when conducting two-sample homogeneity tests.

\begin{table}[!h]
\small
\renewcommand\arraystretch{0.8}
\caption{$p$-values of five homogeneity tests for comparing income distributions between the NE and W groups}
\label{real_example}
\centering
\begin{tabular}{c r r r r r r r r r r}
\toprule
 & Our ELR & $t$-test & Wilcoxon & KS & Cai's ELR\\
\midrule
$p$-value & $7.321\times 10^{-6}$ & 0.0526 & 0.0912 & 0.1657 & 0.0521\\
\bottomrule
\end{tabular}
\end{table}

\section{Concluding Remarks}
\label{Sec_conclusion}
In this paper, we study the problem of testing the homogeneity of two population distributions when the outcomes are subject to nonignorable nonresponse with callback data. We adopt a semiparametric framework that incorporates callback information through Alho's model in \eqref{Alho} and characterises the relationship between the two populations via the DRM in \eqref{DRM}. Within this framework, we propose an ELR statistic for testing homogeneity, which enjoys a Wilks-type property under the null hypothesis, along with an EM algorithm for parameter estimation and test implementation. Simulation studies and real data analysis demonstrate that the proposed test maintains correct Type I error control and achieves competitive power compared with conventional homogeneity tests in the presence of nonignorable nonresponse.

The proposed framework can be extended in several directions. First, the current paper focuses on testing homogeneity between two populations, whereas applications such as the NHIS data often involve more than two groups. Extending the proposed method to accommodate multiple populations is a natural and important direction for future research. Another issue of practical importance concerns potential model misspecification. The proposed procedure relies on the callback model to characterise the response mechanism, and its performance may be affected if this model is misspecified. Developing sensitivity analysis or robustness diagnostics for the proposed homogeneity test under misspecified callback models would therefore be a valuable topic for future investigation.

\section*{Acknowledgements}
This work was supported by the Humanities and Social Sciences Foundation of the Ministry of Education of China (24YJA910005) for C.~Wang; Singapore Ministry of Education Academic Research Tier 1 Fund (A-8000413-00-00) for T.~Yu; and Natural Sciences and Engineering Research Council of Canada (RGPIN-2026-04406) for P.~Li.

\section*{Disclosure statement}

The authors report there are no competing interests to declare.

%

\bibliographystyle{tfnlm}
\bibliography{callback.bib}

\end{document}